\documentclass[aps,pre,showpacs,showkeys]{revtex4-1}
\pdfoutput=1
\usepackage{amssymb}
\usepackage{amsmath}
\usepackage{graphicx}
\usepackage{hyperref}
\newtheorem{theorem}{Theorem}

\newcommand{\figurewidth}{0.77\columnwidth} 

\newcommand*{\re}{\mathrm{e}}            
\newcommand*{\ri}{\mathrm{i}}            

\let\ds\displaystyle
\renewcommand*{\vec}[1]{\mathbf{#1}} 
\renewcommand*{\vector}[1]{\left\langle #1 \right\rangle} 
\let\average\overline
\newcommand*{\grad}{\nabla}
\newcommand*{\Ham}{\ensuremath{\mathcal{H}}} 
\newcommand*{\JJ}{\ensuremath{\mathcal{J}}} 
\newcommand*\mass{M} 
\newcommand*\unit{m} 
\newcommand*\spine{\sigma} 
\newcommand*\residue{n} 
\newcommand*{\EE}{\ensuremath{\mathcal{E}}} 
\newcommand*{\PP}{\mathcal{P}} 
\newcommand*{\ev}{\psi} 
\newcommand*{\vecq}{\vec{q}}
\newcommand*{\vecp}{\vec{p}}

\newcommand*{\vecy}{\vec{y}}

\newcommand*{\half}{\frac{1}{2}}
\newcommand*{\Ai}{\textrm{Ai}}
\newcommand*{\Qy}{Q}
\newcommand*{\vecQy}{\vec{Q}}

\begin{document}

\title{Energetic Pulses in Exciton-Phonon Molecular Chains, and Conservative Numerical Methods for Quasi-linear Hamiltonian Systems}
\author{Brenton~LeMesurier}
\affiliation{Department of Mathematics, College of Charleston, South Carolina}
\altaffiliation{Currently visiting CSCAMM, University of Maryland, and the Department of Mathematical Sciences, George Mason University}
\email{lemesurierb@cofc.edu}
\homepage{http://blogs.cofc.edu/lemesurierb/}
\date{\today}
%
%
%

\begin{abstract}
The phenomenon of coherent energetic pulse propagation in exciton-phonon molecular chains such as $\alpha$-helix protein is studied using an ODE system model of Davydov-Scott type, both with numerical studies using a new unconditionally stable fourth order accurate energy-momentum conserving time discretization, and with analytical explanation of the main numerical observations.

Physically natural impulsive initial data associated with the energy released by ATP hydrolysis are used, and the best current estimates of physical parameter values.
In contrast to previous studies based on a proposed long wave approximation by the nonlinear Schr\"odinger (NLS) equation and focusing on initial data resembling the soliton solutions of that equation,
the results here instead lead to approximation by the third derivative nonlinear Schr\"odinger equation, giving a far better fit to observed behavior.
A good part of the behavior is indeed explained well by the linear part of that equation, the Airy PDE, while other significant features do not fit any PDE approximation, but are instead explained well by a linearized analysis of the ODE system.

A convenient method is described for construction the highly stable, accurate conservative time discretizations used, with proof of its desirable properties for a large class of Hamiltonian systems, including a variety of molecular models.
\end{abstract}

\maketitle

\section{Introduction}

\emph{Exciton-phonon} systems of ODEs are used to model a variety of molecules in which mobile quantum excitations are present along with mechanical degrees of freedom.
A.~Davydov
\cite{Davydov:1971,Davydov+Kislukha:1973} introduced such a model to study energy propagation in $\alpha$-helix protein, present for example in the myocins, kenesins and actin involved in muscular contraction, in chains up to 2000 residues long.
A modified version of Davydov's original equations, is used here, incorporating changes suggested by A.~Scott \cite{Scott:1984} and by Davydov and A.~Zolotariuk in \cite{Davydov+Zolotariuk:1984}:
\begin{equation} \label{Davydov-Scott excitons}
\ri \hbar \frac{d\ev_{\residue}}{dt}
- E_0 \ev_{\residue}
+ J (\ev_{\residue-3}+\ev_{\residue+3})
- L (\ev_{\residue-1}+\ev_{\residue+1})
= \chi (q_{\residue+3}-q_{\residue}) \ev_{\residue}.
\end{equation}
\begin{equation} \label{anharmonic Davydov-Scott phonons}
\mass \frac{d^2 q_{\residue}}{dt^2}
= V'(q_{\residue+3} - q_{\residue}) - V'(q_{\residue} - q_{\residue-3})
+ \chi( |\ev_{\residue}|^2 - |\ev_{\residue-3}|^2),
\end{equation}
This will be called the \emph{Anharmonic Davydov-Scott system}.

Related exciton-phonon systems arise in other molecular models, such as the system
\begin{equation} \label{exciton-phonon chain: excitons}
\ri \hbar \frac{d\ev_{\residue}}{dt}
+ J (\ev_{\residue-1}+\ev_{\residue+1})
= \chi (q_{\residue+1}-q_{\residue-1}) \ev_{\residue},
\end{equation}
\begin{equation} \label{exciton-phonon chain: phonons}
\mass \frac{d^2 q_{\residue}}{dt^2}
=
V'(q_{\residue+1} - q_{\residue}) - V'(q_{\residue} - q_{\residue-1})
+ \chi( |\ev_{\residue+1}|^2 - |\ev_{\residue-1}|^2),
\end{equation}
used to model the conducting polymer polydiacetylene in \cite{Brizhik+al:2000}.
This differs in having two-sided (symmetrical) form of the coupling, and only nearest neighbor interactions, but as should become clear below, the results herein adapt easily to differences such as these.

We will consider in particular pulses in the exciton variables $\ev_n$ that are generated by initial excitation at one end of the chain.
It will be seen that the phenomena are well modeled by a \emph{subsonic limit} leading to a \emph{Helically Coupled Discrete Nonlinear Schr\"odinger equation} [HDNLS]
\begin{equation} \label{HDNLS}
\ri \frac{d\ev_{\residue}}{dt}
+ \hat{J} (\ev_{\residue-3}+\ev_{\residue+3})
- \hat{L} (\ev_{\residue-1}+\ev_{\residue+1})
+ 2 \kappa |\ev_{\residue}|^2 \ev_{\residue}
= 0.
\end{equation}
Further, an important part (but not all) of the pulse propagation can be described with a new long wave PDE approximation; not the nonlinear Schr\"odinger [NLS] model previously proposed by Davydov and considered in numerous subsequent papers, but a \emph{third derivative NLS equation}
\begin{equation} \label{3rd derivative NLS}
\frac{\partial \ev}{\partial t} + \frac{\partial^3 \ev}{\partial x^3} + 2 \ri \kappa |\ev|^2 \ev = 0
\end{equation}
also seen in related work of D.~Pelinowsky and V.~Rothos \cite{Pelinovsky+Rothos:2005}.

Section~\ref{section: Davydov-Scott model} introduces the various mathematical models and their Hamiltonian structures, symmetries and conserved quantities, explaining the successive approximations involved.
Section~\ref{section: Numerical Methods} introduces the accurate, energy and momentum conserving numerical methods used; these are hopefully useful for a wide variety of similar Hamiltonian systems, due to advantages over the symplectic methods often used for such systems.
Section~\ref{section: Numerical Results} presents numerical results, including demonstration of the high degree of accuracy of the successive model simplifications, and the inapplicability (for the present choices of initial data) of the NLS approximations used in various previous studies.
Section~\ref{section: analysis} gives an analytical explanation for many of the phenomena observed, and ends by proposing some ideas for further study.

\section{Modeling Exciton Propagation in $\alpha$-helix Protein and Other Polymers}
\label{section: Davydov-Scott model}

\subsection{The Anharmonic Davydov-Scott ODE System}

The primary mathematical model used here is the above Anharmonic Davydov-Scott system of ODEs (\ref{Davydov-Scott excitons},\ref{anharmonic Davydov-Scott phonons}), which modifies Davydov's original ODE model of $\alpha$-helix protein by adopting a one-sided form for the exciton-phonon coupling (proposed by A.~Scott \cite{Scott:1984} based on the observations of V.~Kuprievich and V.~Kudritskaya \cite{K+K:1982})
and using a nonlinear force for the hydrogen bonds (as introduced by A.~Davydov and A.~Zolotariuk in \cite{Davydov+Zolotariuk:1984}, and resembling the familiar FPU model).
The helical structure of this protein has roughly three residues per twist, with hydrogen bonds  connecting third-nearest neighbors into nearly straight \emph{spines}: spatial proximity leads to attractive exciton coupling along spines in addition to repulsive coupling between neighbors along the molecular backbone as the two dominant exciton interactions.

\textbf{Aside:} many previous publications group residues into unit cells of three residues labelled $\unit$, with the residues within each unit cell labelled by a spine index $\spine = 1, 2, 3$,
but here a single index is more convenient, with explicit third-nearest neighbor interactions.

The variables and parameters in this system are as follows.
\begin{itemize}
\item
Index $n$ labels amino acid residues.
\item
The \emph{exciton} variables $\ev_\residue$ give the probability of excitation of the amide-I mode
in residue $\residue$,
governed by a second quantization Hamiltonian
\[
\Ham_{ex} = E_0 \ev_{\residue}^* \ev_{\residue} 
- J (\ev_{\residue}^* \ev_{\residue+3} + \ev_{\residue+3}^* \ev_{\residue})
+ L (\ev_{\residue}^* \ev_{\residue+1} + \ev_{\residue+1}^* \ev_{\residue}),
\]
where $J$ measures the (attractive) interaction between excitons in residues that are adjacent along a spine and
$L$ measures the (repulsive) interaction between excitons in residues that are adjacent along the molecular backbone.
\item
The \emph{phonon} variables $q_\residue$ are the displacements of the residues from rest position
in the direction of the axis of the helix (that is, along spines), with momenta $p_n = M\dot{q}_n$:
these are associated to the phonon Hamiltonian
\[
\Ham_{ph} = \sum_{\residue} \frac{p_{\residue}^2}{2\mass}
+ V(q_{\residue+3} - q_{\residue})
\]
with $\mass$ the effective mass of each amino acid residue
and
$V(r)$ a potential modeling the hydrogen bond force.
\item
Parameter $\chi$ measures the effect of bond-stretching on the excitons through interaction Hamiltonian
\[
\Ham_{int} = \chi (q_{\residue+3} - q_{\residue}) \ev_{\residue}^* \ev_{\residue}.
\]
\end{itemize}
In fact the $E_0$ term can be eliminated with the transformation
$\ev_{\residue} \to e^{\ri E_0 t}\ev_{\residue}$,
so this is done from here on.
Also, the only anharmonic potential considered is the cubic
\begin{equation} \label{cubic hydrogen bond potential}
V(r) = \frac{K}{2}r^2 - \frac{\gamma}{3} r^3,
\end{equation}
and in fact it will be demonstrated that for the situation studied herein, it is quite adequate to approximate with the harmonic potential $\ds V(r) = \frac{K}{2}r^2$, $K = V''(0)$, as indeed was done by Davydov originally.
This leads to the original ``harmonic'' version of the Davydov-Scott system as proposed by A.~Scott in \cite{Scott:1984}, with phonon equation
\begin{equation} \label{harmonic Davydov-Scott phonons}
\mass \frac{d^2 q_{\residue}}{dt^2}
= K (q_{\residue-3} - 2q_{\residue} + q_{\residue+3}) + \chi( |\ev_{\residue}|^2 - |\ev_{\residue-3}|^2).
\end{equation}

Either form of the system is Hamiltonian, with
$\Ham = \Ham_{ex} + \Ham_{ph} + \Ham_{int}$ and
\begin{equation} \label{complex Hamiltonian equations}
\ri \hbar \frac{d \ev_{\residue}}{dt} = \frac{\partial \Ham}{\partial \ev_{\residue}^*},
\quad
\ri \hbar \frac{d \ev_{\residue}^*}{dt} = - \frac{\partial \Ham}{\partial \ev_{\residue}},
\end{equation}
\begin{equation} \label{real Hamiltonian equations}
\frac{d q_{\residue}}{dt} = \frac{\partial \Ham}{\partial p_{\residue}},
\quad
\frac{d p_{\residue}}{dt} = -\frac{\partial \Ham}{\partial q_{\residue}}.
\end{equation}

\paragraph*{Parameter Values.}
As the results herein are quite robust under variations in the parameter values within the likely range for $\alpha$-helix protein, it is for the most part sufficient to use the values reported in \cite{Scott:1982a,Scott:1984}, which facilitates comparisons to numerous other publications that use those values.
The exciton couplings are best expressed through the frequencies
\[
\hat{J} = J/\hbar \approx \textrm{1.47 THz},
\;
\hat{L} = J/\hbar \approx \textrm{2.33 THz}.
\]
The linear stiffness of the hydrogen bond is
$K \approx \textrm{13 N/m}$.
The effective residue mass $\mass$ is less precisely known, due in part to potential dependence on the particular sequence of amino acids, but it is sufficient to use the typical value
$M \approx \textrm{0.127 zg}$,
which leads to a typical phonon frequency
\[
\omega_0 = \sqrt{K/M} \approx \textrm{10.1 THz},
\]
because it will be seen that the only importance here is that this frequency is substantially larger than the above exciton frequencies.
This puts us in the \emph{subsonic} regime: exciton pulses travel at distinctly lower speeds than the phonons.
As a further consequence, it will be seen in Section~\ref{section: Numerical Results} that the
\emph{subsonic limit} $M \to 0$ (so also $\omega_0 \to \infty$) gives the above HDNLS equation \eqref{HDNLS}, and this approximation is seen in numerical studies to be highly accurate for any physically relevant value of $M$.

Variation of the interaction coefficient $\chi$ has more significant effects, and despite the precise computed value of 34\,pN cited by \cite{Scott:1982a} and various subsequent papers, there is still substantial uncertainty as to its value: the best current estimate appears to be the broad range of experimental values $\chi \approx \textrm{35\,--\,62 pN}$, with computed values subject to far greater uncertainty, even as to its sign \cite{Freedman+al:2010}.
Thus the effect of varying this parameter will be studied:
fortunately, it will be seen that the results herein depend only mildly on this value, with even the linearization $\chi=0$ giving useful information.

\paragraph*{Boundary Conditions.}
The boundary conditions at the ends of the chain depend on if and how the helix is connected to other parts of the molecule, but here the simplest, unbound form is assumed:
``out of bounds'' values of $\ev_n$ and of the \emph{bond-stretchings} $r_n := q_{\residue+3} - q_{\residue}$ are effectively neglected in the Hamiltonian so for such index values
\begin{equation}
\label{HDBC's}
\ev_n = 0, \, r_n = 0.
\end{equation}
For constructing simplified PDE models via a long wave approximation,
it is also convenient to consider an infinite chain with $\residue \in \mathbb{N}$ and
$\ev_n \to 0, \; r_n \to 0$ as $|n| \to \infty$.

\paragraph*{Initial Data.}
The initial data considered will be the physically plausible cases for an initial excitation caused by the energy release in ATP hydrolysis: primarily initial excitation at one residue.
The most interesting phenomena will be seen to arise from excitation at one end of the chain, so
\begin{equation} \label{single end impulse}
\ev_1(0) = 1, \ev_n(0) = 0 \mbox{ for } n > 1.
\end{equation}
ATP hydrolysis can also excite a pair of neighboring residues, so there will be brief comments on the variant $\ev_1(0) = \ev_2(0) = 1$.
An initially still chain is used:
$q_n(0) = p_n(0) = 0.$

\subsection{Momenta (conserved quantities other than the Hamiltonian)}

The equations above have a conserved \emph{exciton number}
$\EE = \sum_n \ev_{\residue}^* \ev_{\residue}.$
This is related to the probability density of quantum mechanics, but as noted above, it need not be unity, due to the possibility of multiple initial excitations.
This invariant is associated via Noether's Theorem with a linear symmetry group action,
the gauge symmetry
\begin{equation} \label{gauge symmetry}
\ev_{\residue} \to \re^{\ri s}\ev_{\residue}, \quad \ev_{\residue}^* \to \re^{-\ri s}\ev_{\residue}^*.
\end{equation}

The Davydov-Scott system also has a conserved momentum $\PP_\spine$ on each spine:
$\PP_\spine = \sum_{\unit} p_{3\unit+\sigma}$
associated with spine translation symmetries $q_{3\unit+\sigma} \to q_{3\unit+\sigma} + s_\spine$.
However, conservation of linear momentum is respected by almost any reasonable time discretization (for example, any Runge-Kutta method) so no more will be said about this.

\subsection{Approximation by a Helically Coupled Nonlinear Schr\"odinger Equation}

The Davydov-Scott system has several disparate scales in both space and time, and these can be used to derive simpler approximations.
The first is that for physically relevant initial data, it will be seen in the numerical results of Section~\ref{section: Numerical Results} that the bond-stretchings $r_n$ are of small amplitude so that the linearized force $-K r$ is an adequate approximation, corresponding to harmonic potential $V(r) =\frac{K}{2}r^2$.
Next is the \emph{subsonic limit} approximation:
the frequency $\omega_0 = \sqrt{K/M}$ is considerably higher than the exciton coupling frequencies $\hat{J}$ and $\hat{L}$, and in practice exciton phenomena are on an even slower scale, so that variation in the amplitude $|\ev_n|$ is far slower that that of the mechanical variables $q_n$.
For small $M$, solving Eq.~\eqref{harmonic Davydov-Scott phonons} by variation of parameters gives
\[
r_\residue = q_{\residue+3}-q_{\residue} = - \frac{\chi}{K} |\ev_{\residue}|^2
+ \text{oscillations of characteristic frequency $\omega_0$}
\]
and it is plausible that the excitons respond primarily to the slowly varying moving average part,
which is given by setting $\mass = 0$ in \eqref{harmonic Davydov-Scott phonons}.
Using this moving average approximation
\begin{equation}
r_\residue \approx - \frac{\chi}{K} |\ev_{\residue}|^2
\end{equation}
in the exciton equation \eqref{Davydov-Scott excitons} eliminates the mechanical variables, reducing the model to the Helically Coupled Discrete Nonlinear Schr\"odinger [HDNLS] equation \eqref{HDNLS}, with
\[
\kappa := \frac{\chi^2}{2 \hbar K} \approx \textrm{0.45 -- 1.4 THz}.
\]
This has Hamiltonian
\begin{equation}
\Ham =
\sum_n
- J (\ev_{\residue}^* \ev_{\residue+3} + \ev_{\residue+3}^* \ev_{\residue})
+ L (\ev_{\residue}^* \ev_{\residue+1} + \ev_{\residue+1}^* \ev_{\residue})
- \kappa (\psi_n^*\psi_n)^2. 
\end{equation}
The validity of this approximation is demonstrated numerically in Section~\ref{section: Numerical Results} below.

\section{Energy-Momentum Conserving Time Discretizations}
\label{section: Numerical Methods}

To study these systems and assess the adequacy of the above HDNLS approximation,
some numerical solutions should be considered.
For that, the necessary numerical methods will now be described, and this is done for a general Hamiltonian system
\begin{equation} \label{general Hamiltonian system}
\frac{d \vecy}{dt}
 = \JJ \grad_\vecy \Ham(\vecy)
 = \JJ \frac{\partial \Ham}{\partial \vecy}(\vecy)
\end{equation}
with $\JJ$ an anti-symmetric matrix.


\paragraph*{Notation.}
We will focus on the time advance map for single time step,
from a time $t$ to $t + \delta t$.
For any scalar variable $y$ or vector $\vec{y}$,
we use the variable's name alone to denote its value at time $t$,
$t^+ = t + \delta t$,
$y^+ = y(t^+) = y(t + \delta t)$,
$\delta y = y^+ - y$,
and
$\average{y} = \ds\frac{y+y^+}{2}$.

\subsection{Discrete Gradient Methods for Exact Energy Conservation}

Exact conservation of invariants has been seen to be a desirable feature of numerical methods for Hamiltonian systems; see for example \cite{Geometric_Numerical_Integration:2006}.
Following ideas originating in the work of
O.~Gonzalez and
J.~Simo
\cite{Gonzales:1996,Gonzales+Simo:1996},
the first step is to ensure conservation of the Hamiltonian (energy) by approximating such a system
by a \emph{discrete Hamiltonian system}
\begin{equation} \label{discrete Hamiltonian system}
\frac{\delta \vecy}{\delta t} = \JJ \tilde{\grad}_\vecy \Ham (\vecy,\vecy^+)
\end{equation}
using a suitable \emph{discrete gradient} approximation
\begin{equation} \label{discrete grad f}
\tilde{\grad}_\vecy f (\vecy,\vecy^+) \approx \grad_\vecy f (\vecy).
\end{equation}
that satisfies the \emph{Discrete Chain Rule}
\begin{equation} \label{DMCR}
\delta f
= (\tilde{\grad}_\vecy f)(\vecy,\vecy^+) \cdot \delta \vecy.
\end{equation}
This condition is assumed from now on, along with linearity and the consistency condition
\begin{equation} \label{DG-G consistency}
\lim_{\vecy^+ \to \vecy} (\tilde{\grad}_\vecy f) (\vecy,\vecy^+) = \grad_\vecy f(\vecy).
\end{equation}
Component notation like
$\tilde{\grad}_\vecy f (\vecy,\vecy^+) = \vector{\tilde{D}_{y_1} f (\vecy,\vecy^+), \dots}$
will occasionally be used.

Conservation of energy is easily shown for a discrete gradient method by mimicking the familiar argument for (continuous) Hamiltonian systems: using in succession \eqref{DMCR}, \eqref{discrete Hamiltonian system}, and the anti-symmetry of $\JJ$,
\begin{equation*}
\frac{\delta \Ham}{\delta t}
= (\tilde{\grad}_\vecy \Ham) (\vecy,\vecy^+) \cdot \frac{\delta \vecy}{\delta t}
= (\tilde{\grad}_\vecy \Ham) (\vecy,\vecy^+) \cdot \JJ (\tilde{\grad}_\vecy \Ham) (\vecy,\vecy^+)
= 0.
\end{equation*}

\subsection{Choosing a Discrete Gradient that Also Respects Quadratic Momenta}
\label{subsection: momentum conserving DG}

Many such ``energy conserving'' discrete gradients can be found, but conserving other invariants (here all called \emph{momenta}) requires an appropriate choice of the gradient approximation.
It will be seen that there is a natural limitation to quadratic (including linear) momenta, but this is sufficient for most systems of physical relevance.
Here the approach introduced in
\cite{LeMesurier:2010,LeMesurier:2011}
is followed, based on three facts:
\begin{enumerate}
\item
There is a unique discrete gradient for functions of a single variable $y$
\begin{equation} \label{df/dy}
\tilde{\grad}_y f(y,y^+) := \left\{ \begin{array}{ll}
\ds \frac{\delta f}{\delta y}, & y^+ \neq y
\\[2ex]
\ds \frac{df}{dy}(y), & y^+ = y
\end{array}\right.
\end{equation}
following from the chain rule requirement \eqref{DMCR}.
For polynomials, this simplifies in a way that avoids the division by zero issue, via
\begin{equation} \label{discrete power law}
\tilde{\grad}_y y^{p+1} = y^n + y^{n-1}y^+ \cdots + (y^+)^n.
\end{equation}
\item
There is a unique time-reversal symmetric discrete gradient for a product of two variables
\begin{equation} \label{delta (yj yk)}
\delta  (y_j y_k) = \average y_j \delta y_k +  \average y_k \delta y_j 
\end{equation}
which corresponds to evaluating the true gradient at the midpoint:
\begin{equation} \label{grad (yj yk)}
\tilde{\grad} (y_j y_k) (\vecy, \vecy^+) = \grad (y_j y_k) (\average\vecy).
\end{equation}
In fact this extends to a \emph{discrete product rule} based on
\begin{equation} \label{discrete product rule}
\delta (f g) =  \average f \delta g +  \average g \delta f. 
\end{equation}

Thus linear terms in the equations, corresponding to quadratic terms in the Hamiltonian, are discretized exactly as for the implicit midpoint rule, which is a popular momentum conserving symplectic method for Hamiltonian systems.
The only differences are for nonlinearities, which for the systems of interest herein are those coming from the Hamiltonian terms
\begin{equation} \label{hamiltonian nonlinearities}
\chi (q_{\residue+3} - q_{\residue}) \ev_{\residue}^* \ev_{\residue},\; \frac{\gamma}{3}(q_{n+3}-q_n)^3
\; \text{for Eq's~(\ref{Davydov-Scott excitons},\ref{anharmonic Davydov-Scott phonons}) },\;
\kappa(\psi_n^*\psi_n)^2\; \text{for Eq.~\eqref{HDNLS}.}
\end{equation}

\item
Many physically relevant Hamiltonian systems with conserved momenta have a natural form in which all the momenta are quadratic (including linear) functions of the state variables, and are related through Noether's theorem to a group of affine symmetries of the Hamiltonian $\Ham$, with invariance of $\Ham$ manifested by the fact that it can be expressed as a composition
\begin{equation} \label{H-Hhat}
\Ham(\vecy) = \hat{\Ham}(\vecQy),
\end{equation}
where each component $\Qy_m$ of the new state vector
$\vecQy$
is a quadratic
\begin{equation} \label{Qy}
\Qy_m = \half \sum_{j,k} A_{m}^{jk} y_j y_k + \sum_j b_m^j y_j,
\; \mbox{(each $A_m := \{ A_{m}^{jk} \}$ symmetric)}
\end{equation}
that is invariant under the symmetry group.
For example, with the systems seen herein, the invariant quadratics with which the Hamiltonian can be expressed are the exciton products
$e_{n,m} = \ev_n^*\ev_m$ and the bond-stretchings $r_\residue$.
In particular, the nonlinear terms seen here are
\begin{equation} \label{hamiltonian nonlinearities factorized}
\chi r_{\residue} e_{\residue,\residue},\, \frac{\gamma}{3} r_\residue^3, \text{ and } \kappa e_{\residue,\residue}^2. 
\end{equation}

The discrete Jacobian of this change of variables is given by the true Jacobian evaluated at the midpoint:
\[
\tilde{D}_\vecy \vecQy (\vecy, \vecy^+) = {D}_\vecy \vecQy (\average\vecy).
\]
\end{enumerate}

These facts and the above chain rule requirement naturally lead to:
\begin{equation} \label{DG of H=Hhat(Qy)}
\tilde{\grad}_\vecy \Ham
= \sum_m \tilde{D}_m \hat{\Ham}(\vecQy,\vecQy^+) \; \tilde\grad_\vecy \Qy_m(\vecy,\vecy^+),
= \sum_m \tilde{D}_m \hat{\Ham} (\vecQy,\vecQy^+) \; \grad_\vecy \Qy_m (\average \vecy).
\end{equation}
For the nonlinearities herein, the discrete gradients are now determined by the factorizations in \eqref{hamiltonian nonlinearities factorized} through simple forms:
\begin{equation} \label{DG's for DS and HDNLS}
\tilde{D}_{r}(r e) = \average e,
\tilde{D}_{e}(r e) = \average r,
\tilde{D}_{e}(e^2) = 2 \average e,
\tilde{D}_{r} (r^3) =r^2 +r r^+ + (r^+)^2,
\end{equation}
using \eqref{discrete power law} for the last.

Using such a discrete gradient, energy and momenta will be conserved with any choice for the factors
$\tilde{D}_m \hat{\Ham} (\vecQy,\vecQy^+)$.
In practice, the above rules for single variable functions, products, compositions, and linearity are generally enough to construct a suitable discrete gradient for $\hat\Ham$.

\begin{theorem} \label{theorem 1}
For a Hamiltonian system
\begin{equation*}
\frac{d \vecy}{dt} = \JJ \grad_\vecy \Ham(\vecy),
\quad
\Ham(\vecy) = \hat{\Ham}(\vecQy)
\end{equation*}
as described above, and thus with a discrete gradient
\[
\tilde{\grad}_\vecy \Ham =
\sum_m \tilde{D}_m \hat{\Ham} (\vecQy,\vecQy^+) \; \grad_\vecy \Qy_m (\average \vecy),
\]
solving numerically
by the corresponding discrete gradient method
\begin{equation} \label{discrete gradient method}
\frac{\vecy^+ - \vecy}{\delta t}
= \JJ \sum_m \tilde{D}_m \hat{\Ham} (\vecQy,\vecQy^+) \;
\grad_\vecy \Qy_m \left( \frac{\vecy + \vecy^+}{2} \right)
\end{equation}
conserves the Hamiltonian and all the quadratic momenta.
\end{theorem}

\noindent\textbf{Proof of Theorem 1} 
Energy conservation is already established above, so consider conservation of an invariant $Q$.
Such quadratics are in fact invariant for any Hamiltonian $\Ham = \hat\Ham(\vecQy)$ constructed from the quadratic forms $Q_m$ as in \eqref{H-Hhat}, including the alternative choices $\Ham_m := Q_m$,
and invariance of $Q$ on each of those Hamiltonian flows means that
\begin{equation} \label{dQ/dt=0}
0 = \frac{d Q}{d t}
= \grad Q \cdot \JJ \grad \Ham_m
= \grad Q \cdot \JJ \grad \Qy_m,
\end{equation}
so that we have vanishing of the Poisson brackets
\begin{equation} \label{PB Q Q_m = 0}
\{ Q, \Qy_m \}(\vecy) :=  \grad Q(\vecy) \cdot \JJ \grad \Qy_m(\vecy) = 0.
\end{equation}

Mimicking \eqref{dQ/dt=0} for the discrete flow and using the fact from \eqref{grad (yj yk)} that discrete gradients of quadratics are given by the true gradients at the midpoint gives
\begin{equation*}
\label{delta Q = 0}
\frac{\delta Q}{\delta t}
= \tilde\grad Q \cdot \JJ \tilde\grad \Ham
= \grad Q(\average\vecy) \cdot \JJ \sum_m (\tilde{D}_m \hat{\Ham}) \grad \Qy_m(\average\vecy)
= \sum_m (\tilde{D}_m \hat{\Ham}(\vecQy, \vecQy^+))\{ Q,\Qy_m \}\kern-0.3ex(\average\vecy).
\end{equation*}
Evaluating \eqref{PB Q Q_m = 0} at $\vecy = \average\vecy$ gives
$\{ Q,\Qy_m \}\kern-0.3ex(\average\vecy) = 0$, so \mbox{$\delta Q = 0$}.

\subsection{Practical Implementation: an Iterative Solution Method}

The system of equations will be nonlinear (unless the Hamiltonian system itself is linear),
so we need an iterative solution method.
To exploit the quasi-linearity of the system to preserve linear stability properties and exact momentum conservation without the cost of a full quasi-Newton method, we proceed as follows:
construct successive approximations $\vecy^{(k)}$ of $\vecy^+$ by solving
\begin{equation}
\label{iterative conservative DG scheme}
\frac{\vecy^{(k)} - \vecy}{\delta t}
= \JJ \sum_m \tilde{D}_m \hat{\Ham} (\vecQy,\vecQy^{(k-1)}) \;
\grad_\vecy \Qy_m (\average{\vecy}^{(k)}),
\end{equation}
where
$\average{\vecy}^{(k)} = (\vecy + \vecy^{(k)})/2$
and
$\vecQy^{(k-1)} = \vecQy(\vecy^{(k-1)})$,
and initialization can be with $\vecy^{(0)} = \vecy$ or some other suitable approximation of $\vecy^+$.

That is, the nonlinear part $\tilde{\grad}_{\vecQy} \hat{\Ham}$ is approximated using the current best available approximation $\vecy^{(k-1)}$ of $\vecy^+$,
while the linear terms are left in terms of the unknown $\vecy^{(k)}$ to be solved for.
This equation is linear in the unknown $\vecy^{(k)}$, making its solution straightforward, particularly with the narrow coupling bandwidth of the coupling in the systems studied here.
Much as above, we have:
\begin{theorem} \label{theorem 2}
Each iterate $\vecy^{(k)}$ given by the above scheme \eqref{iterative conservative DG scheme}
conserves all quadratic first integrals that are conserved by the original discrete gradient scheme
\eqref{discrete gradient method}.
\end{theorem}
The proof is as for Theorem~\ref{theorem 1} except that the Poisson brackets are evaluated at
\mbox{$\left( \vecy + \vecy^{(k)} \right)/2$}.

This approach to iterative solution also gives unconditional linear stability, since as noted above,
for a linear system it is the same as the A-stable implicit midpoint method, and indeed only a single iteration is needed in that case.
Energy is of course only conserved in the limit $k \to \infty$, but iterating until energy is accurate within machine rounding error is typically practical:
if this take too many iterations, it is better for overall accuracy to reduce the time step size $\delta t$ to speed the convergence.

\subsection{Time Discretization for the Davydov-Scott System}

Applying the above results to the anharmonic Davydov-Scott system is mostly a matter of separating linear  terms from nonlinear, applying the implicit midpoint rule to the former and using
Eq.~\eqref{DG's for DS and HDNLS} in Eq.~\eqref{iterative conservative DG scheme} for the latter:
\begin{equation}
\ri \frac{\delta \ev_n^{(k)}}{\delta t}
+ \hat{J} \left( \average{\ev}_{\residue-3}^{(k)} + \average{\ev}_{\residue+3}^{(k)} \right)
- \hat{L} \left( \average{\ev}_{\residue-1} + \average{\ev}_{\residue+1} \right)
= \frac\chi\hbar \left( \average{q}_{\residue+3}^{(k-1)} - \average{q}_{\residue}^{(k-1)} \right) \average{\ev}_{\residue}^{(k)},
\end{equation}
\begin{equation}
\mass \frac{\delta q_{\residue}^{(k)}}{\delta t} = \average{p}_{\residue}^{(k)},
\end{equation}
\begin{equation}
\begin{split}
\frac{\delta p_{\residue}^{(k)}}{\delta t}
= & K (\average{q}_{\residue+3}^{(k)} - \average{q}_{\residue}^{(k)})
+ \chi \left( \average{|\ev_{\residue}^{(k-1)}|^2} - \average{|\ev_{\residue-3}^{(k-)}|^2} \right)
\\
&- \frac{\gamma}{3}
\bigg[ \left( q_{\residue+3}^{(k-1)} - q_{\residue}^{(k-1)} \right)^2
 + \left( q_{\residue+3}^{(k-1)} - q_{\residue}^{(k-1)} \right)
    \left( (q^+)_{\residue+3}^{(k-1)} - (q^+)_{\residue}^{(k-1)} \right)
\\
&+ \left( (q^+)_{\residue+3}^{(k-1)} - (q^+)_{\residue}^{(k-1)} \right)^2 \bigg].
\end{split}
\end{equation}

\subsection{Higher Order Accuracy by Symmetric Step Composition}
The methods seen so far are only second order accurate in time.
Fortunately, the method of \emph{symmetric step composition},
(developed by
M.~Creutz, A.~Gocksch, E.~Forest, M.~Suzuki and H.~Yoshida
\cite{Creutz+Gocksch:1989, Forest:1989, Suzuki:1990, Yoshida:1990}
for use with symplectic methods, and reviewed
by E.~Hairer, C.~Lubich, and G.~Wanner
in the book
\cite{Geometric_Numerical_Integration:2006})
gives a systematic way to construct methods of any higher even order while preserving all the interesting properties:
conservation of the Hamiltonian and quadratic invariants, time-reversal symmetry, and unconditional stability.

Numerical results are computed below by combining the above discrete gradient method with the fourth-order accurate Suzuki form of step composition
\cite[Example II.4.3, p.~45]{Geometric_Numerical_Integration:2006}:
compose five discrete gradient steps of lengths
$\rho_j \delta$,
\[
\rho_1 = \rho_2 = \rho_4 = \rho_5 = \frac{1}{4 -\sqrt[3]{4}} \approx 0.41, 
\quad
\rho_3 = 1 - 4\rho_1 \approx -0.66. 
\]

\subsection{Comparisons to Other Methods}

The most commonly used conservative methods for Hamiltonian systems are symplectic methods, which can conserve momenta but cannot in general conserve energy, as described by a theorem of Z.~Ge and J.~Marsden \cite{Ge+Marsden:1988}.
In the present situation with stiff systems of ODEs and Hamiltonian not of purely mechanical form
$\Ham(\vecq,\vecp) = K(\vecp) + U(\vecq)$,
the preferred choices of symplectic method are the implicit midpoint method, higher order Diagonally Implicit Runge-Kutta [DIRK] methods, and fully implicit Gaussian Runge-Kutta methods.

All DIRK symplectic methods are cognates of the energy-momentum methods described here, given by applying the same step composition procedures to the implicit midpoint method instead of to the discrete gradient method.
It has been illustrated in \cite{LeMesurier:2010,LeMesurier:2011} that the basic discrete gradient method can handle qualitative features of solutions better than the midpoint method, though this has not been tested directly when step composition is applied to each method.

Gaussian symplectic methods can be desirable when the time step size is small enough to allow their solution by simple fixed point iteration, but are not cost effective for stiff systems, where an unconditionally stable iterative method such as that above is highly desirable.

\section{Numerical Results}
\label{section: Numerical Results}

As the initial excitation due to ATP hydrolysis will occur at at most two residues, the initial state is very far from the slowly varying form assumed in long wave approximations by PDE's.
Thus one question addressed here, as in earlier work like \cite{Scott:1982a,Scott:1984},
is whether solutions with such initial data evolve into a form that can be well-approximated at later times by a smooth function of position, leading to a hopefully more tractable PDE model.

\paragraph*{Time Step Choice.}
The choice of time steps here is always cautiously constrained by
\[
\delta t \leq \min\left(\frac{1}{2(\hat{J} + \hat{L})}, \frac{1}{\omega_0}\right),
\]
which satisfies the natural accuracy and stability requirements for explicit methods, and for convergence of basic fixed point iterative solution of the nonlinear schemes.
However it is confirmed that accurate solutions, in the sense that all graphs of exciton data are completely indistinguishable from results with smaller time steps, are given for any time step size
\[
\delta t \leq \frac{1}{2(\hat{J} + \hat{L})}
\]
depending only on the time scale manifested in the exciton evolution equation.
Thus the time discretization is effectively handling any faster time scales in the mechanical variables in the innocuous way that one hopes for stiff modes to be handled by an unconditionally stable method, with no adverse effect on the accuracy of the more slowly evolving (exciton) variables.

\subsection{Numerical Observations for the Davydov-Scott and HDNLS Systems}

\begin{figure}[htbp]
\begin{center}
\includegraphics[width=\figurewidth]{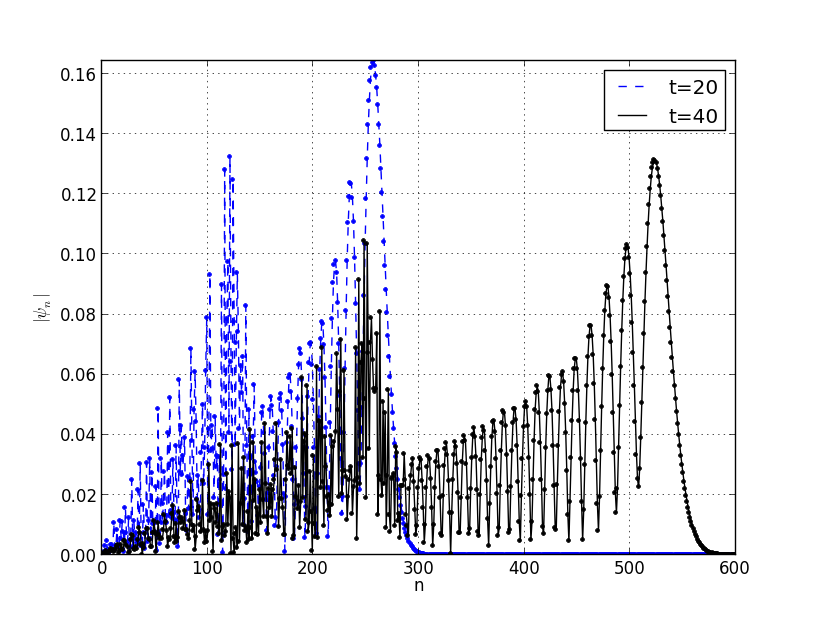}
\caption{Anharmonic Davydov-Scott system, $\gamma = 4$, $\chi=35$:
$|\ev_n|^2$ at times $t = 20,40$.}
\label{fig1}
\end{center}
\end{figure}

We first solve the Anharmonic Davydov-Scott system
(\ref{Davydov-Scott excitons},\ref{anharmonic Davydov-Scott phonons})
with 1000 residues,
hydrogen bond nonlinearity of cubic form \eqref{cubic hydrogen bond potential} with $\gamma = 4$,
and initial excitation at one end as in \eqref{single end impulse}.
Figure~\ref{fig1} is for $\chi=35$, the minimum of the likely range cited above, showing the exciton amplitude $|\ev_n|$ at times $t=20$ and $40$.
It reveals a dominant leading pulse of speed about 13.3 residues per unit time that is slowly varying in $n$, and a secondary pulse of speed about 6.4 with no slow spatial variation.

\begin{figure}[htbp]
\begin{center}
\includegraphics[width=\figurewidth]{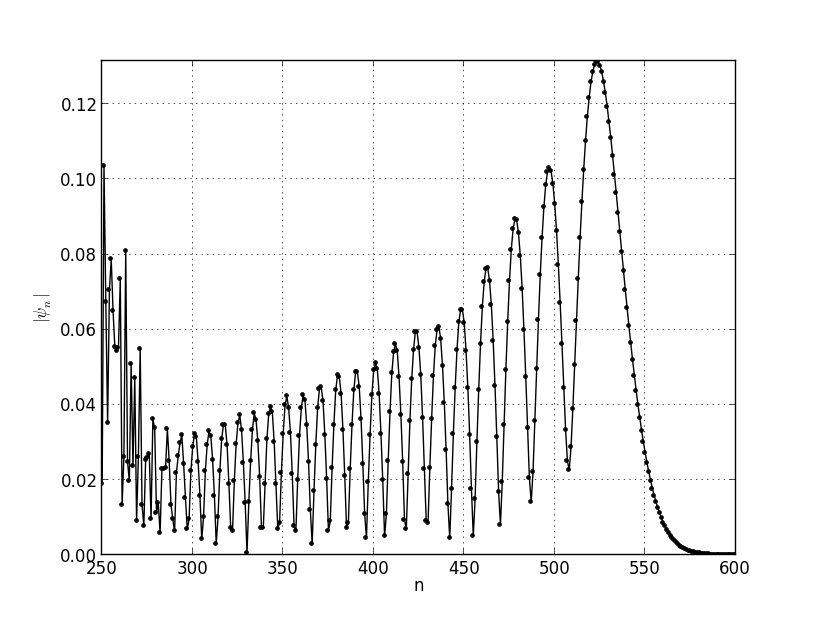}
\caption{Anharmonic Davydov-Scott system, as in Figure~\ref{fig1} except at $t = 40$ only: $|\ev_n|^2$.}
\label{fig2}
\end{center}
\end{figure}

\begin{figure}[htbp]
\begin{center}
\includegraphics[width=\figurewidth]{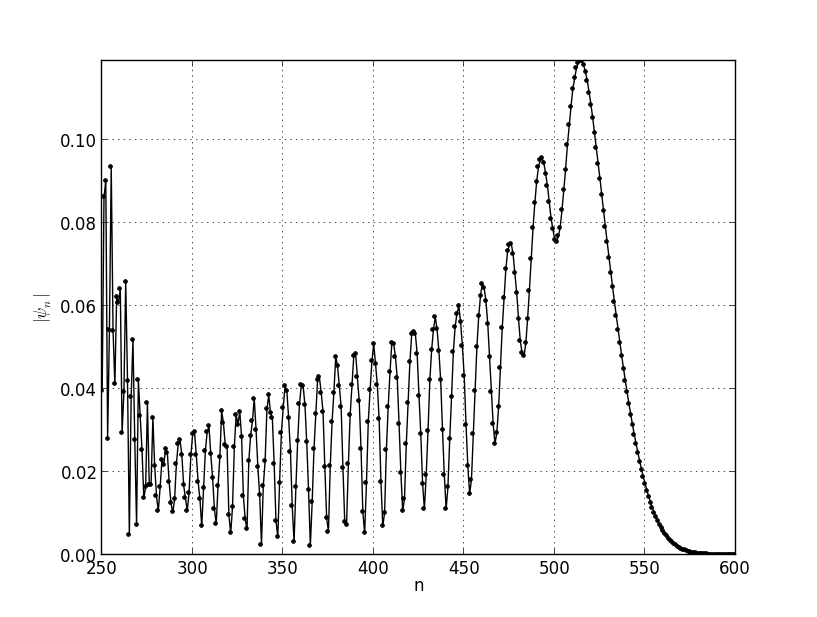}
\caption{Anharmonic Davydov-Scott system, as in Figure~\ref{fig2} except with $\chi=62$: $|\ev_n|^2$.}
\label{fig3}
\end{center}
\end{figure}

The time evolution is very similar in all cases, so it is sufficient to compare at a single time $t=40$ from now on.
Figure~\ref{fig2} repeats the above data at that time, and Figure~\ref{fig3} is the same except for $\chi=62$, the other extreme of the likely range of values.
Although a significant quantitative difference is seen, the qualitative description above still holds for the stronger nonlinearity, and it will be seen soon that other key features are also unchanged.
(The latter is also similar to what is seen of \cite{Scott:1982a},
which however used the two-point initial impulse form $\ev_1(0) = \ev_2(0) = 1$, $\ev_n(0) = 0$ for $n>2$, and $\chi=34$.)

\begin{figure}[htbp]
\begin{center}
\includegraphics[width=\figurewidth]{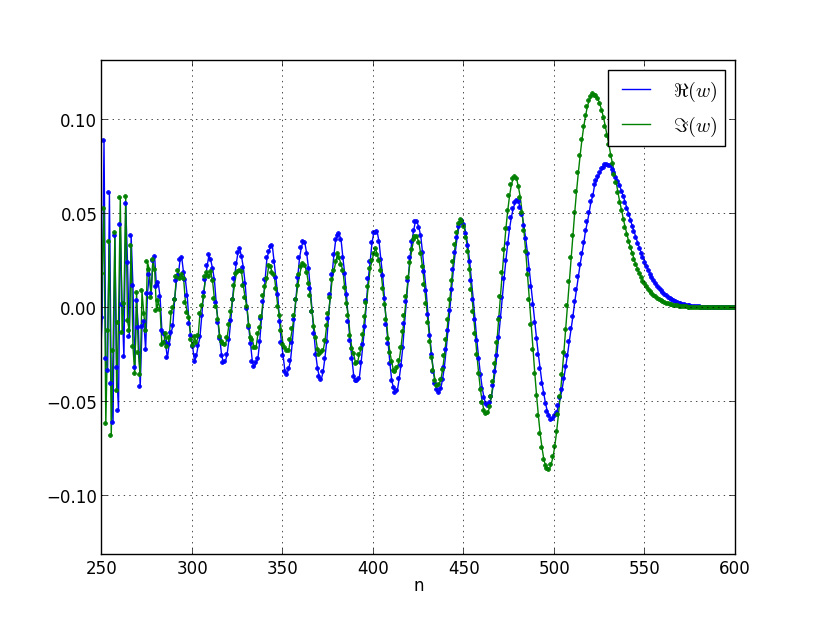}
\caption{Real and imaginary parts of $w_n = \ri^n \ev_n$ for $\chi=35$.}
\label{fig4}
\end{center}
\end{figure}

The slow variation of exciton amplitude suggests the possibility of a long wave PDE approximation for this part of the solution, as proposed by Davydov and others.
However, slow variation is not seen in $\ev_n$ as a whole, due to rapid phase variation, and this is true even if one restricts to individual spines.
Instead, the phase advances by a factor of approximately $-\ri$ at each step along the chain, and thus by  factor of $\ri$ at each step along a spine.
This is best revealed by studying $w_\residue := \ri^\residue \ev_\residue$: the real and imaginary parts of this are shown in Figures~\ref{fig3} and \ref{fig4} for the two cases above.

\begin{figure}[htbp]
\begin{center}
\includegraphics[width=\figurewidth]{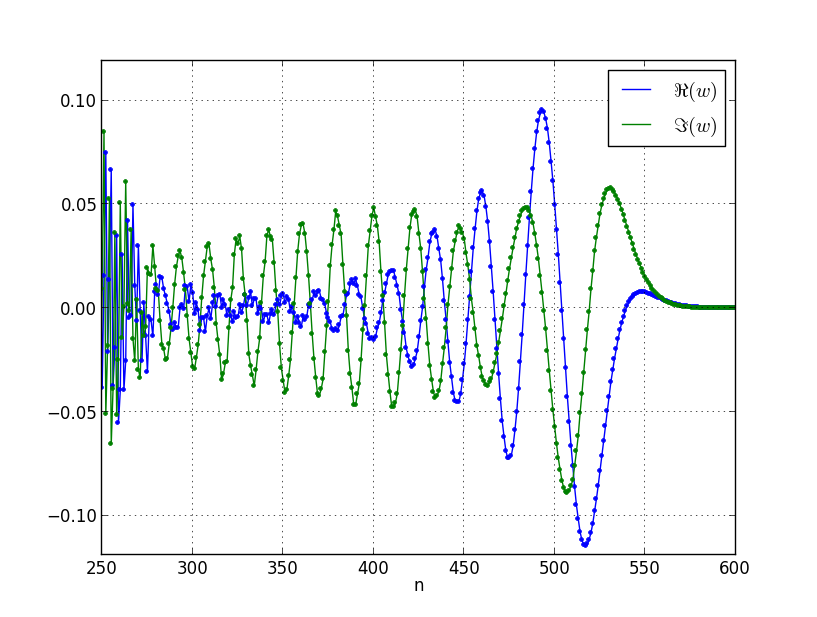}
\caption{Real and imaginary parts of $w_n = \ri^n \ev_n$ for $\chi=62$.}
\label{fig5}
\end{center}
\end{figure}

\begin{figure}[htbp]
\begin{center}
\includegraphics[width=\figurewidth]{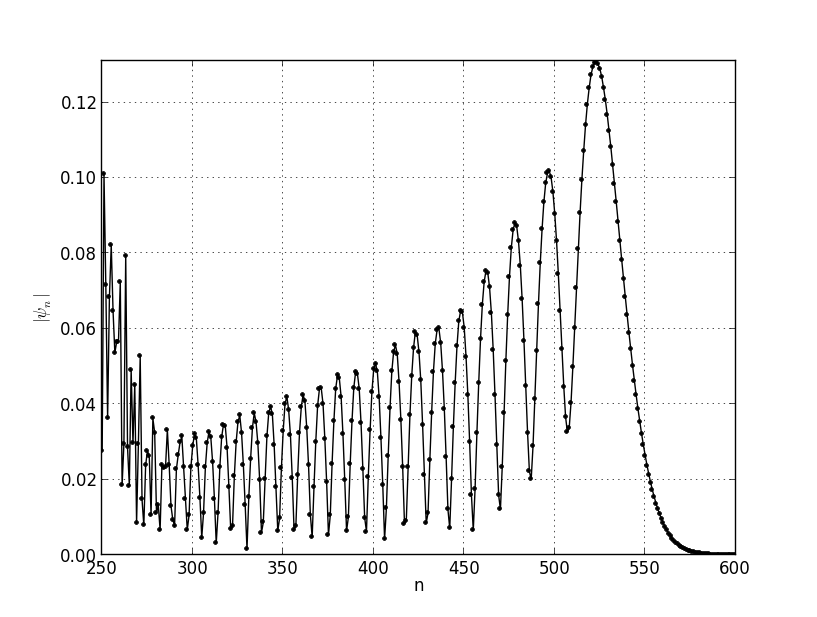}
\caption{Harmonic Davydov-Scott system, $\chi=35$: $|\ev_n|^2$.}
\label{fig6}
\end{center}
\end{figure}

Next, it can observed that the nonlinearity of the hydrogen bonds is of little significance, due to the magnitude of $r_\residue$ staying quite small: less than about $0.3$.
This is indicated by Figure~\ref{fig6} for the harmonic case $\gamma=0$, with $\chi=35$.

\begin{figure}[htbp]
\begin{center}
\includegraphics[width=\figurewidth]{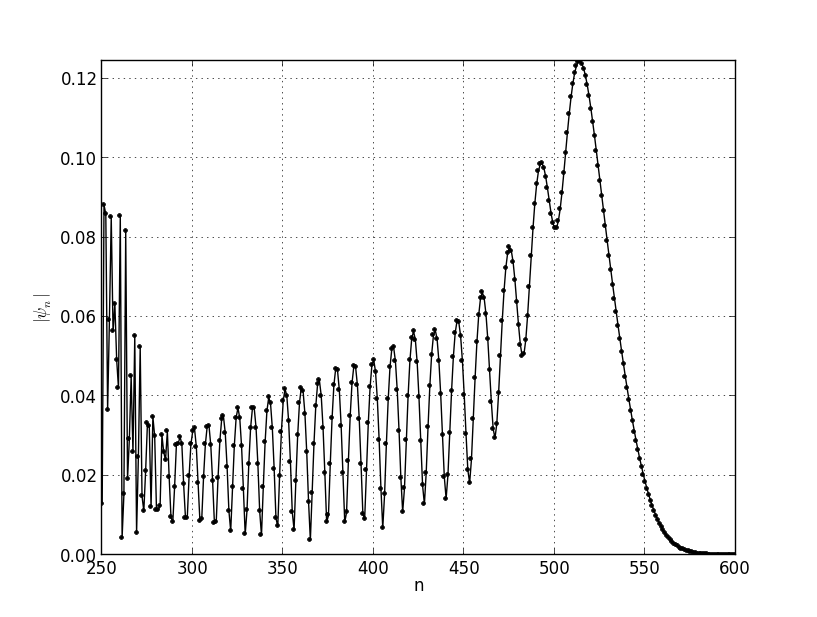}
\caption{HDNLS system, $\chi=62$: $|\ev_n|^2$.}
\label{fig7}
\end{center}
\end{figure}

However, this point is made more emphatically by considering the next level of approximation, by the subsonic limit of HDNLS \eqref{HDNLS}.
Even for the harder case of $\chi=62$, the exciton form is little changed, as seen in Figure~\ref{fig7}, and it is much the same over the full range of likely $\chi$ values.

\subsection{The Linear Approximation $\chi \to 0$}

\begin{figure}[htbp]
\begin{center}
\includegraphics[width=\figurewidth]{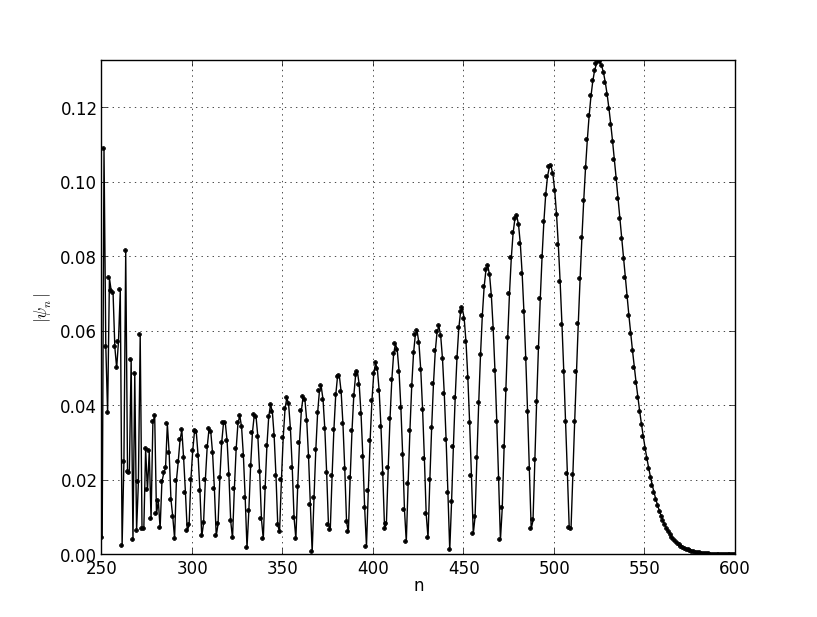}
\caption{Linearization $\chi=0$: $|\ev_n|^2$.}
\label{fig8}
\end{center}
\end{figure}

\begin{figure}[htbp]
\begin{center}
\includegraphics[width=\figurewidth]{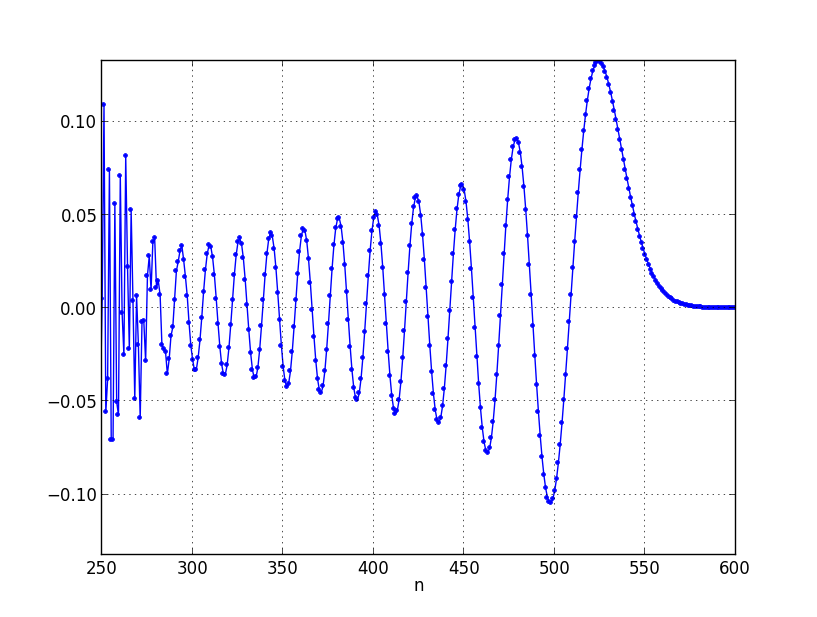}
\caption{Linearization $\chi=0$: $w_n$, which is now real-valued.}
\label{fig9}
\end{center}
\end{figure}

A final approximation worth considering is $\chi \to 0$,
which for either the Davydov-Scott or HDNLS systems gives a linear equation for the excitons alone:
\begin{equation} \label{HDSE}
\ri \frac{d\ev_{n}}{dt} + \hat{J} (\ev_{n+3}+\ev_{n-3}) - \hat{L} (\ev_{n+1}+\ev_{n-1}) = 0.
\end{equation}
This will be the starting point for the analysis below, but first it can be observed that at least some main qualitative features of the above solutions are retained in this linear model, as seen in Figures~\ref{fig8} and \ref{fig9}.
The form of $w_n$ might now be recognized as resembling the Airy function $\Ai$, and this will be explained in the analysis of the next section.

\subsection{Brief Remarks on Other Cases}

Some brief observations for other choices of initial data and parameter values.
\begin{enumerate}
\item
For an initial impulse at other locations, one has exciton self-trapping, with most of the signal staying at the initial location. There are weaker pulses propagating in each direction, which are well explained by the analysis of linearized equations given in the next section.
\item
For larger values of $\chi$, about 100 and up, there is again strong exciton self-trapping, with little signal propagation.
\item
For the double excitation initial data $\ev_1(0) = \ev_2(0) = 1$, $\ev_n(0) = 0$ for $n > 2$ as considered in \cite{Scott:1982a}, the behavior is similar to that discussed here, though with somewhat stronger nonlinear effects.
\item
For a simplified exciton-phonon model such as in
equations~(\ref{exciton-phonon chain: excitons},\ref{exciton-phonon chain: phonons}) and/or with the symmetric coupling form seen in \eqref{exciton-phonon chain: phonons}, the main phenomena are similar, though with the pulse velocity of course changed to $2\hat{J}$.
\end{enumerate}

\section{Analysis, and the Third Derivative NLS Approximation}
\label{section: analysis}

Previous studies have proposed a PDE approximation based on the assumption that $\ev_\residue(t)$ varies slowly in $n$, leading to PDEs related to the nonlinear Schr\"odinger equation, and thus to the study of solutions related to its traveling wave solutions of hyperbolic secant form.
However, it is seen above that for the impulsive initial data considered herein, $\ev_\residue$ does not become slowly varying in phase.
Instead, slow variation along the chain is seen in the transformed quantity $w_\residue = \ri^\residue \ev_\residue$, for which the Davydov-Scott exciton evolution equation
\eqref{anharmonic Davydov-Scott phonons} becomes
\begin{equation} \label{Davydov-Scott excitons w}
\frac{dw_{\residue}}{dt}
+ \hat{J} (w_{\residue+3} - w_{\residue-3})
+ \hat{L} (w_{\residue+1} - w_{\residue-1})
= -\ri \frac{\chi}{\hbar} (q_{\residue+3}-q_{\residue}) w_{\residue},
\end{equation}
and HDNLS becomes
\begin{equation} \label{HDNLS w}
\frac{dw_{\residue}}{dt}
+ \hat{J} (w_{\residue+3} - w_{\residue-3})
+ \hat{L} (w_{\residue+1} - w_{\residue-1})
= 2 \ri \kappa |w_{\residue}|^2 w_{\residue}.
\end{equation}
Recalling that $\hat{J} \approx \textrm{1.47 THz}$ and $\hat{L} =  \textrm{2.33 THz}$ whereas $\kappa \approx \textrm{0.45 -- 1.4 THz}$, and that our initial data ensures $|w_\residue| \leq 1$ with far smaller values typical, it appears likely that the linearization
\begin{equation} \label{HDLS w}
\frac{dw_{\residue}}{dt}
+ \hat{J} (w_{\residue+3} - w_{\residue-3})
+ \hat{L} (w_{\residue+1} - w_{\residue-1})
= 0
\end{equation}
is a useful first approximation.
One initial observation is that for the initial data considered herein, this has real valued solutions, fitting with the observed phase behavior.

Following the approach of D.~Pelinowsky and V.~Rothos \cite{Pelinovsky+Rothos:2005}, we seek solutions of the form
\begin{equation}
\ev_n = \re^{\ri (\beta n + \omega t)} w(\tau, z), \quad z = n-vt, \tau = \epsilon t, \epsilon \ll 1
\end{equation}
where the fast spatial and time scales are isolated in an exponential factor, leaving a slowly varying envelope $w(\tau, z)$.
In the limit $\chi \to 0$, these should relate to ``discrete traveling wave'' solutions of the linearization \eqref{HDSE}, with
\begin{equation}
w(\tau, z) = \re^{\ri k z}, \quad |k| \ll 1.
\end{equation}
This has dispersion relation
\begin{equation}
\label{dispersion relation} 
\omega(k) = kv + 2 \hat{J} \cos(3(\beta + k)) - 2 \hat{L} \cos(\beta + k),
\end{equation}
and thus group velocity
\begin{equation}
v = 6 \hat{J} \sin(3\beta) - 2 \hat{L} \sin(\beta),
\end{equation}
with maximum
\begin{equation}
v = v_{max} = 6 \hat{J} + 2 \hat{L} \approx 13.48
\end{equation}
occurring for $\beta = -\pi/2$,
$\omega = 0$,
so that
\begin{equation}
\ev_n = (-\ri)^n w(z),
\end{equation}
the same transformation suggested above based on numerical observations.
(There is a left going counterpart of course, excluded by the initial data used here.)

One way to see this is that from initially impulsive initial data with a wide range of wave numbers present, there is a clustering of signals of various wave numbers at critical numbers of group velocity $dv/d\beta = 0$, in particular at
$\beta = -\pi/2$, which gives the maximum velocity.
There are in fact six critical numbers, with the other two that correspond to right-going pulses forming a supplementary pair $\beta' \approx 0.15\pi$, $\beta'' = \pi - \beta'$ with the same velocity $v' \approx 6.60$, fitting well the velocity of about $6.4$ observed for the second slower pulse above.
This double root allows pulses with spatial dependence given by the real and imaginary parts $\sin(\beta' n)$, $\cos(\beta' n)$ which explains the break-down of slow amplitude variation seen for that second pulse.

Returning to the Davydov-Scott system, we now seek solutions similar to this.
Nonlinearity requires an amplitude scaling, so we use
\begin{equation} \label{continuum approximation}
w_n(t) \approx \sqrt\epsilon \, u(z, \tau),
\end{equation}
where
$v = 6\hat{J}+2\hat{L}$,
$\tau = \epsilon t$,
$\epsilon = (1/3)a k^3$,
$a = (27\hat{J} + \hat{L})$.

This gives
\begin{equation}
w_\tau + w_{zzz} = \ri \hat\chi q_z w,
\end{equation}
and in the subsonic limit of HDNLS,
\begin{equation}
w_\tau + w_{zzz} + \ri  \hat\chi |w|^2 w = 0,
\end{equation}
which is sometimes called the \emph{third derivative nonlinear Schr\"odinger equation}.

For either of these equations, the linearization is the Airy PDE
\begin{equation}
w_\tau + w_{zzz} = 0,
\end{equation}
and the impulsive initial conditions considered here can be associated with its fundamental solution
\begin{equation}
w(z,\tau) = \frac{1}{(3 \tau)^{(1/3)}} \Ai \left(\frac{z}{(3 \tau)^{(1/3)}}\right).
\end{equation}

Converting back gives approximate solution
\begin{equation}
w_n(t) \approx  \frac{\sqrt{k} \, a^{1/6}}{\sqrt{3} \, t^{1/3}}
 \Ai \left(\frac{n - v t}{(a t)^{1/3}}\right).
\end{equation}

\paragraph*{Proposals for further analysis.}
For the related case of discrete NLS equations of the form
\begin{equation}
\ri \frac{d\ev_n}{dt} + \ev_{n-1} + \ev_{n+1}
+ \chi f( \ev_{n-1}, \ev_n, \ev_{n+1})
\end{equation}
with cubic nonlinearities $f$ having the gauge symmetry \eqref{gauge symmetry}, Pelinowsky and Rothos \cite{Pelinovsky+Rothos:2005} showed that solutions of the nonlinear equation for small $\chi$ bifurcate from solutions of the linearization at certain points, in particular the one $k=0$, $\omega=0$, $\beta = -\pi/2$ seen above.
It seems likely that a similar analysis would apply here.
%
Beyond that, what the numerical results suggest, and which should be analyzed further, is that the nonlinearity provides some ``dispersion management'', preventing the leading pulse from spreading as fast as in the linearization, and making it more dominant compared to the following oscillation train.

\section{Conclusions}

\begin{enumerate}
\item
The sustained traveling exciton pulses seen in Davydov-style exciton-oscillator models
of energy propagation in $\alpha$-helix protein
are well approximated by the subsonic, small mass approximation, giving a variant of the discrete NLS equation.
\item
As noted by other authors, the main part of the pulse has magnitude $|\ev_n|$ that varies slowly, suggesting a long wave PDE approximation.
However, the phase of the $\ev_n$ varies rapidly in index $n$, by about a quarter turn at each step, and thus the slow spatial variation is instead in $w_n = \ri^n \ev_n$.
This leads to a new PDE approximation by the third derivative nonlinear Schr\"odinger equation
$w_\tau + w_{zzz} = \ri |w|^2 w$,
which indeed gives solutions fitting well to the fastest moving part of the solutions.
\item
Linearization of this to the Airy PDE $w_\tau + w_{zzz} = 0$ also gives a good qualitative fit to many features such as pulse speed, with the main nonlinear deviation being in the most intense front-most part of the pulse.
\item
Analysis of the linearized discrete system also explains a good part of the observed behavior: it is not nearly as accurate as the above nonlinear PDE in describing the leading part of the pulse, but explains the second, slower pulse for which the PDE is not applicable.
\item
Evidence of nonlinear ``self-trapping'' effects are seen, in that the leading hump of the pulse remains stronger and narrower as time increases than those of the linearization, supporting more sustained propagation than a linear model would predict.
\item
The higher order exactly energy-momentum conserving time-discretization method used is seen to handle well the stiffness that can arise in such systems, making it a good candidate for similar problems, including spatial discretization of various stiff nonlinear dispersive PDE's.
\end{enumerate}


\begin{thebibliography}{BECHO00}

\bibitem[BECHO00]{Brizhik+al:2000}
Larissa Brizhik, Alexander Eremko, Leonor Cruzeiro-Hansson, and Yulia
  Olkhovska.
\newblock {\em Physical Review B}, 61:1129, 2000.

\bibitem[CG89]{Creutz+Gocksch:1989}
M.~Creutz and A.~Gocksch.
\newblock Higher order hybrid {Monte-Carlo} algorithms.
\newblock {\em Phys. Rev. Lett.}, 63:9--12, 1989.

\bibitem[Dav71]{Davydov:1971}
Alexandr~S. Davydov.
\newblock {\em Theory of Molecular Excitations}.
\newblock Plenum press, New York, 1971.

\bibitem[DK73]{Davydov+Kislukha:1973}
Alexandr~S. Davydov and N.~I. Kislukha.
\newblock Solitary excitations in one-dimensional molecular chains.
\newblock {\em Phys. Status Solidi B}, 59:465--470, 1973.

\bibitem[DZ84]{Davydov+Zolotariuk:1984}
Alexandr~S. Davydov and A.~V. Zolotariuk.
\newblock {\em Physica Scripta}, 30:426, 1984.

\bibitem[FMC10]{Freedman+al:2010}
Holly Freedman, Paulo Martel, and Leonor Cruzeiro.
\newblock Mixed quantum-classical dynamics of an {amide-I} vibrational
  excitation in a protein alpha-helix.
\newblock {\em Phys. Rev. B}, 82(17):174308, 2010.

\bibitem[For89]{Forest:1989}
E.~Forest.
\newblock Canonical integrators as tracking codes.
\newblock {\em AIP Conference Proceedings}, 184:1106--1136, 1989.

\bibitem[GM88]{Ge+Marsden:1988}
Z.~Ge and J.~E. Marsden.
\newblock {Lie-Poisson Hamilton-Jacobi} theory and {Lie-Poisson} integrators.
\newblock {\em Phys. Lett. A}, 133:134--139, 1988.

\bibitem[Gon96]{Gonzales:1996}
O.~Gonzales.
\newblock Time integration and discrete {Hamiltonian} systems.
\newblock {\em Journal of Nonlinear Science}, 6:449--467, 1996.

\bibitem[GS96]{Gonzales+Simo:1996}
O.~Gonzales and Juan~C. Simo.
\newblock On the stability of symplectic and energy-momentum algorithms for
  nonlinear {Hamiltonian} systems with symmetry.
\newblock {\em Comput. Methods Appl. Mech. Eng.}, 134:197--222, 1996.

\bibitem[HLW06]{Geometric_Numerical_Integration:2006}
Ernst Hairer, Christian Lubich, and Gerhard Wanner.
\newblock {\em Geometric Numerical Integration: Structure Preserving Algorithms
  for Ordinary Differential Equations}.
\newblock Springer, 2nd edition, 2006.

\bibitem[KK82]{K+K:1982}
V.~A. Kuprievich and V.~Kudritskaya.
\newblock Preprint ITP-82-64E, Institute for Theoretical Physics, Kiev, 1982.

\bibitem[LeM12a]{LeMesurier:2011}
B.~LeMesurier.
\newblock Conservative unconditionally stable discretization methods for
  {Hamiltonian} equations, applied to wave motion in lattice equations modeling
  protein molecules.
\newblock {\em Physica D}, 241(1):1--10, January 2012.
\newblock Published online 1 Oct 2011.

\bibitem[LeM12b]{LeMesurier:2010}
Brenton LeMesurier.
\newblock Studying {Davydov's} {ODE} model of wave motion in alpha-helix
  protein using exactly energy-momentum conserving discretizations for
  {Hamiltonian} systems.
\newblock {\em Mathematics and Computers in Simulation}, 82(7):1239--1248,
  2012.
\newblock Published online 30 December 2010.

\bibitem[PR05]{Pelinovsky+Rothos:2005}
D.~Pelinovsky and V.~Rothos.
\newblock Bifurcations of travelling wave solutions in the discrete {NLS}
  equation.
\newblock {\em Physica D}, 202:16--36, 2005.

\bibitem[Sco82]{Scott:1982a}
Alwyn~C. Scott.
\newblock The vibrational structure of {Davydov} solitons.
\newblock {\em Physica Scripta}, 25:651--658, 1982.

\bibitem[Sco84]{Scott:1984}
Alwyn~C. Scott.
\newblock Launching a {Davydov} soliton: {I.} {Soliton} analysis.
\newblock {\em Physica Scripta}, 29:279--283, 1984.

\bibitem[Suz90]{Suzuki:1990}
M.~Suzuki.
\newblock Fractal decomposition of exponential operators with applications to
  many-body theories and {Monte-Carlo} simulations.
\newblock {\em Phys. Lett. A}, 135:319--323, 1990.

\bibitem[Yos90]{Yoshida:1990}
H.~Yoshida.
\newblock Construction of higher order symplectic integrators.
\newblock {\em Phys. Lett. A}, 150:262--268, 1990.

\end{thebibliography}

\end{document}